\documentclass[showkeys,showpacs,aps,pre,numerical,twocolumn,10pt,superscriptaddress]{revtex4-1}
\usepackage{graphicx}
\usepackage{amsmath,amssymb}
\usepackage{times,mathptmx}
\usepackage{units}
\usepackage{epsfig}
\usepackage{booktabs}
\usepackage{multirow}

\usepackage[latin1]{inputenc}
\usepackage[english]{babel}
\usepackage{color}

\newcommand{\ITP}{Institut f{\"u}r Theoretische Physik,
  Technische Universit{\"a}t Berlin, 10623 Berlin, Germany}
\newcommand{\FU}{Institut f{\"u}r Theoretische Physik,
  Freie Universit{\"a}t Berlin, 14195 Berlin, Germany}

\begin{document}

\date{\today}
\title{Amplitude-phase coupling drives chimera states in globally
coupled laser networks }

\author{Fabian B{\"o}hm}\affiliation{\ITP}
\author{Anna Zakharova}\affiliation{\ITP}
\author{Eckehard Sch{\"o}ll}\affiliation{\ITP}
\author{Kathy L{\"u}dge}\affiliation{\FU}

\begin{abstract}
For a globally coupled network of semiconductor lasers with delayed optical feedback, we demonstrate the existence
of chimera states. The domains of coherence and incoherence that are typical for chimera states are found
to exist for the amplitude, phase, and inversion of the coupled lasers. These chimera states defy several of the
previously established existence criteria. While chimera states in phase oscillators generally demand nonlocal
coupling, large system sizes, and specially prepared initial conditions, we find chimera states that are stable for global coupling in a network of only four coupled lasers for random initial conditions. The existence is linked to a regime of multistability between the synchronous steady state and asynchronous periodic solutions. We show that amplitude-phase coupling, a concept common in different fields, is necessary for the formation of the chimera states.

\end{abstract}


\maketitle

Synchronization is a common phenomenon in interacting nonlinear dynamical systems 
in various fields of research such as physics, chemistry, biology, engineering, or 
socio-economic sciences \cite{HAK83,PIK01,BOC02}. While a lot of knowledge has been gained on the origin of complete synchronization, more complex partial synchronization patterns have only recently become the focus of intense research. We still lack a full understanding of these phenomena, and a very prominent example are chimera states where an ensemble of identical elements self-organizes into spatially separated coexisting domains of coherent (synchronized) and incoherent (desynchronized) dynamics \cite{KUR02a,ABR04}.  Since their first discovery a decade ago many theoretical investigations of coupled phase oscillators and other simplified models have been carried out \cite{MOT10,PAN14}, but their experimental observation in real systems was only reported very recently in optical light modulators \cite{HAG12}, optical comb \cite{VIK14}, chemical \cite{TIN12}, mechanical \cite{MAR13,KAP14}, electronic \cite{LAR13}, and electrochemical \cite{WIC13,SCH14a}
oscillator systems. Theoretical studies have found chimeras also in other systems, 
including higher-dimensional systems \cite{OME12a,PAN13,PAN14}, e.g., 
spiral wave chimeras \cite{SHI04,gu13}, FitzHugh-Nagumo neural 
systems \cite{OME13}, Stuart-Landau oscillators \cite{LAI10,BOR10,SET14}, where pure amplitude chimeras \cite{ZAK14} were found, or quantum interference devices \cite{LAZ14}. 
In real-world systems chimera states might play a role, e.g., in the unihemispheric sleep of birds and dolphins~\cite{RAT00}, in neuronal bump states~\cite{LAI01,SAK06a}, in power grids~\cite{MOT13a}, or in social systems~\cite{GON14}. 

Although no universal mechanism for the formation of chimera states has yet been established, three general essential requirements have been found in many studies: (i) a large number of coupled 
elements, (ii) non-local coupling, and (iii) specific initial conditions. These were primarily derived from the phase oscillator model \cite{PAN14} but also apply to other systems. 
If these conditions are not met, the chimera states tend to have very short lifetimes. Recent studies, however, suggest that these paradigms can be broken and chimera states are observed also for small system sizes \cite{ASH14}, global coupling \cite{BAN14,SET14,SCH14a,YEL14} and random initial conditions \cite{XIE14}. 

Surprisingly, chimera states appear at the interface of independent 
fields of research putting together different scientific communities. 
Recent examples are quantum chimera state \cite{VIE14} or coexistence 
of coherent and incoherent patterns with respect to the modes of an 
optical comb \cite{VIK14,LAR14}. The aim of the present study is to provide a bridge between laser networks and chimera patterns, which opens up numerous perspectives for application of chimera states and at the same time provides more insight into their understanding on the conceptual level. While synchronization phenomena have been well studied for small 
networks of coupled laser systems \cite{HEI01b,WUE05a,HIC11,FLU11b,SOR12,SOR13}, chimera states 
have not yet been reported in laser networks so far. This sparks the 
question what conditions have to be met for the formation of chimera 
states in laser networks. It has been shown in theory and experiments that
amplitude-phase coupling has an important influence upon
the synchronization behaviour. 

Coupled amplitude-phase dynamics is a significant concept not only for laser dynamics. It is a well-established paradigm, which is widely exploited in various fields of research. In nonlinear dynamics it typically refers to anisochronicity, i.e., the dependence of the frequency of oscillations on their amplitude \cite{ZAK10a}. In fluid dynamics a similar principle is known as shear: any real fluids moving along a solid boundary will incur a shear stress at that boundary \cite{KLA10}. In the laser community the term amplitude-phase coupling defines a concept which implies that the phase of the electric field inside the laser cavity is dynamically linked to its amplitude. This is induced by changes in the susceptibility of the 
gain material with the density of carriers.

Recent theoretical \cite{LIN12b,LIN14} and experimental \cite{MEL06,GRI08} investigations have led to a critical re-assessment of this concept and its role in dynamical instabilities, in particular in quantum dot lasers. Here we show that amplitude-phase coupling is an essential driving mechanism for the occurrence of long-living chimera patterns in globally coupled laser networks.
Another concept used in laser dynamics is delayed optical feedback from an external cavity. 
This has prominent applications in controlling the dynamics of lasers \cite{SCH07,FLU13}, however, only few studies have investigated their combined influence
upon partial synchronization and chimera states as done in this Letter.
Networks of mutually coupled semiconductor lasers combine both effects, and are a versatile model 
system for complex network dynamics which also allows for easy experimental realization. Moreover, it has great 
practical relevance for modern communication technologies. 
Our aim is to combine ideas from network science and laser dynamics to gain further 
insight into the formation and existence criteria of chimera states. 
With this, we also address the three common chimera criteria discussed above, and 
show that coupled lasers represent a stunning counterexample, which breaks all three paradigms simultaneously.

\begin{figure}[ht]
    \centering
  \includegraphics[width=0.4\textwidth]{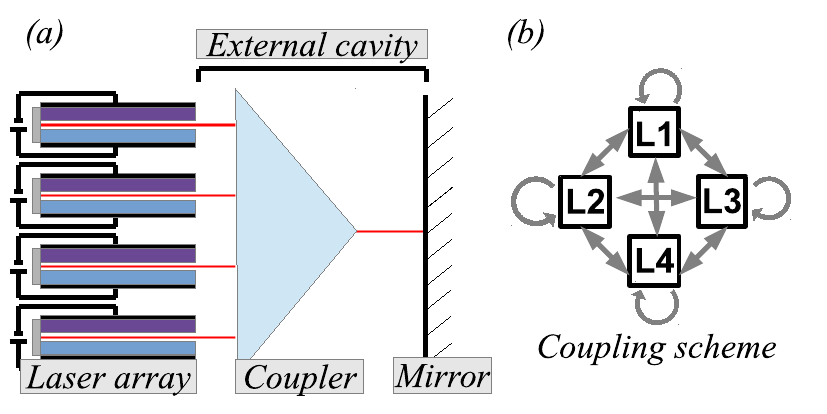}
    \caption{(a) Scheme of a laser array coupled by a common mirror via an external cavity. (b) All-to-all coupling scheme.}
    \label{fig1}
\end{figure}
For our setup we choose an array of $Z$ identical lasers where the output is globally
coupled into a single external cavity (see Fig. \ref{fig1}). The
lasers receive feedback from one common mirror at the end of the
external cavity with a time delay $\tau$ (cavity roundtrip time). We model this system by 
dimensionless semi-classical rate equations, i.e. Lang-Kobayashi 
equations\cite{LAN80b,ROZ75}.
\begin{eqnarray}
    \frac{dE_i}{dt}&=&(1+i\alpha)E_i N_i + e^{iC_p}\kappa 
    \sum_{j}^{Z}e^{iC_p}E_j(t-\tau)
    \label{E}\\
    \frac{dN_i}{dt}&=&\frac{1}{T}(p-N_i-(1+2N_i)|E_i|^2)
    \label{N}
\end{eqnarray}
Equation \eqref{E} governs the complex electric field $E_i$ of the
i-th laser in the array, and Eq. \eqref{N} describes the
inversion $N_i$ of the electrons, $i=1,...,Z$. The amplitude-phase coupling is 
modelled by the linewidth enhancement factor $\alpha$. For 
semiconductor lasers, typical values are $\alpha=2.5$ \cite{ERN10b}. 
$T$ is the ratio of the lifetime of the electrons in the upper level and 
of the phonons in the laser cavity. The lasers are pumped electrically with the excess pump rate $p$. The
feedback strength $\kappa$ and the feedback phase $C_p$ are the
bifurcation parameters, which are used to tune the dynamics and the
synchronization behavior of the system. They are determined from
the reflectivity of the mirror and the length of the external
cavity. Writing the complex electrical field in polar
representation $E_i=A_i(t)e^{i\varphi_i(t)}$ introduces the amplitude
$A_i(t)$ and phase $\varphi_i(t)$. The stationary solutions of the
Lang-Kobayashi equations are given by the external cavity modes (ECM).
Their general form is charaterized by the complex electric field
vector $E_i(t)=A_i(t)e^{i\omega_i t}e^{i\Phi_i}$
rotating with a frequency $\omega_i$, an arbitrary phase
offset $\Phi_i$, and the amplitude $A_i$ in the complex plane. The
inversion $N_i$ is in a steady state.

\begin{figure}[t!]
    \centering
  \includegraphics[width=0.4\textwidth]{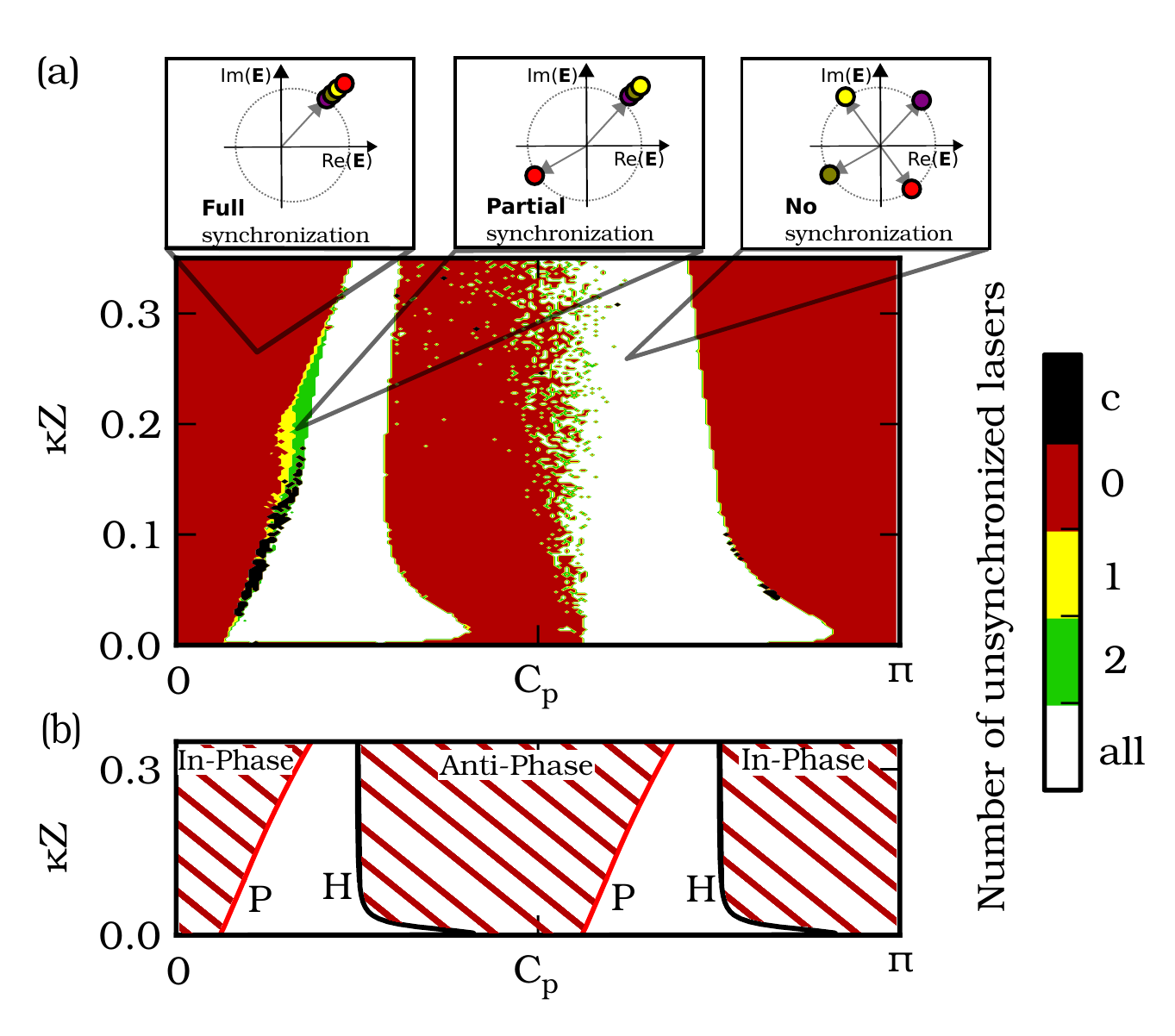}
    \caption{(a) Synchronization regimes in the plane of feedback phase $C_p$ vs. effective feedback strength $\kappa Z$  for 
    a network of $Z=4$ lasers obtained by numerical integration of 
    Eqs.(\ref{E}),(\ref{N}). The color code denotes the number of 
    lasers which synchronize in one group, $c$ denotes cluster 
    synchronization. The insets sketch regimes of 
    full synchronization (left), partial synchronization (center), and 
    no synchronization (right) for the electrical field vectors. (b): Synchronization regimes of two coupled lasers with zero delay obtained by path continuation. The 
synchronous (in- and anti-phase) and asynchronous regimes are  
separated by a pitchfork (P) and a Hopf bifurcation (H) line. 
Parameters: $T=392$,$\alpha=2.5$,$p=0.23$, and (a) $\tau=1$, (b) $\tau=0$.}
    \label{fig2}
\end{figure}

Depending on the amplitude and the phase differences between the 
oscillators, it is possible to realize different forms of
synchronization. The simplest case is full (zero-lag) synchronization, where
$\omega_i=\omega_j$, $\Phi_i-\Phi_j=0$ and $A_i=A_j$ holds for all
lasers $i,j=1,...,Z$ at all times $t$ (see Fig.\ref{fig2}, inset on top left). 
In a partially synchronized state either the frequencies or the
amplitudes are different in one or more lasers (see Fig \ref{fig2}, inset on top center). In an unsynchronized state there is no fixed phase relation between the oscillators 
(see Fig.\ref{fig2}, inset on top right). Fig.\ref{fig2}(a) shows the 
synchronization behaviour of a 4-laser network in the parameter space of feedback phase $C_p$ and 
effective feedback strength $\kappa Z$. For the initial conditions of 
the numerical simulation, the phases of the individual lasers are 
randomly distributed along the complex unit circle. The amplitudes and 
inversions are chosen identical for all lasers: $A_i(0)=1, N_i(0)=0$. 
We observe two distinct domains of desynchronization (white) that 
separate the stable synchronous regimes (red).

To understand the synchronization behaviour of the network in the ($C_p$,$\kappa$)
plane, we first consider the case of full synchronization. Inserting
the ECM solution into \eqref{E} and \eqref{N} with $E_i=E_j$ yields a system which is equivalent 
to two mutually coupled lasers with the coupling strength $\kappa Z$.  
A straigth-forward linear stability analysis of the synchronous state of two coupled
lasers for the case without delay \cite{KOZ00,KOZ01,VLA03,YAN04c} 
yields that the synchronous state destabilizes through a
supercritical Hopf and a pitchfork bifurcation. Utilising the
systems' symmetries, two additional bifurcation lines can be found.
The Lang-Kobayashi equations are invariant under a change of sign 
in $E$, thus a solution shifted by an angle of $\pi$ ($Ee^{i\pi}=-E$) is
also a solution. This gives two additional bifurcation lines which
enclose the anti-phase regime. In this regime, the lasers
organize in phase-locked pairs with identical amplitudes and a phase-shift of $\pi$ between
them.

The bifurcation structure obtained numerically for a 4-laser network from simulation of 
Eqs.(\ref{E}),(\ref{N}) in Fig. \ref{fig2}(a) is compared with the 2-laser model 
and zero delay in panel (b). Fig. \ref{fig2}(b) shows the pitchfork and 
Hopf (H) bifurcation lines of the 2-laser model obtained by path 
continuation. Despite the simplifications, the shapes of the bifurcation lines and the position of the 
in-phase and anti-phase synchronous regimes are very similar to the 
4-laser model with time delay. We have checked by increasing the system 
size $Z$ in the path continuation plot that the difference in the 
shapes of the unsynchronized regions is due to the exclusion of time 
delay.

\begin{figure}[t!]
    \centering 
  \includegraphics[width=0.4\textwidth]{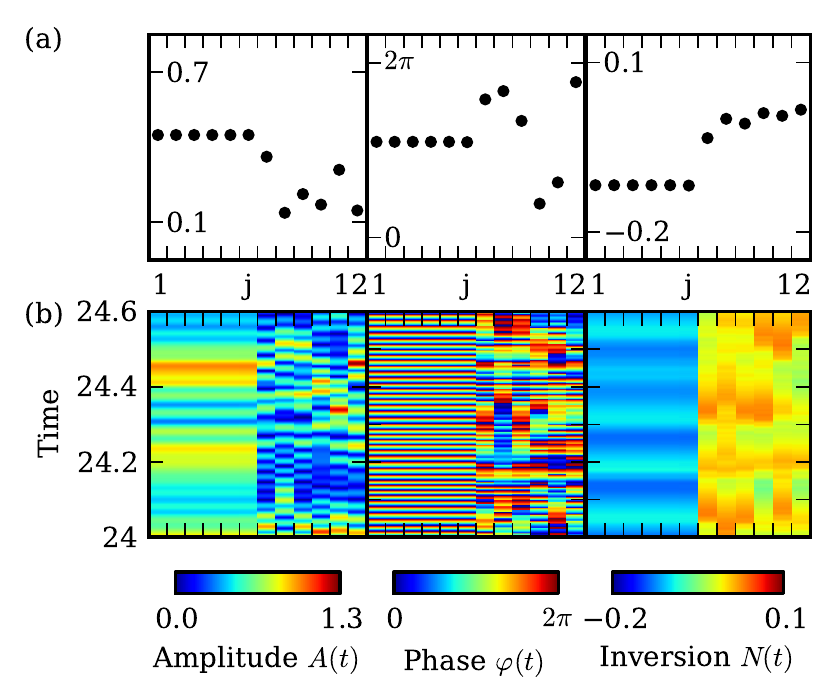}
    \caption{Chimera state in a network of 12 coupled lasers. (a): 
    Snapshots of amplitudes $A_j$, phase $\varphi_j$, inversion $N_j$. (b): Space-time plots of 
    $A_j(t)$, $\varphi_j(t)$, $N_j(t)$.
        Parameters: $\kappa=0.017$, $C_p=0.55$, $T=392$, $\alpha=2.5$, 
        $p=0.23$, $\tau=1$.}
    \label{fig3}
\end{figure}

The dynamics of the 4-laser network with time delay shows continuous 
wave (CW) emission of all lasers inside the in-phase and anti-phase synchronous regimes. Crossing 
the pitchfork and Hopf bifurcation lines, all lasers desynchronize. In the 
desynchronized state amplitude, inversion, and phase become chaotic. While the transition from
synchronization to desynchronization usually occurs suddenly when crossing the bifurcation lines, we find a small regime near the pitchfork bifurcation at
low feedback strengths where the transition occurs gradually. In this 
regime the unsynchronized and the synchronized 
lasers self-organize into domains of coherent and incoherent dynamics and 
form a chimera-like state. A space-time plot of the amplitude, phase, and inversion dynamics for such a state is shown in Fig.\ref{fig3} (b). Because of the global coupling no spatial ordering of the lasers is a priori defined. For better visualisation, the permutation invariance of the laser indices in Eqs.\eqref{E} and \eqref{N} has been exploited, and the lasers have been re-numbered such that coherent and incoherent domains are separated.

\begin{figure}[t!]
    \centering
  \includegraphics[width=0.4\textwidth]{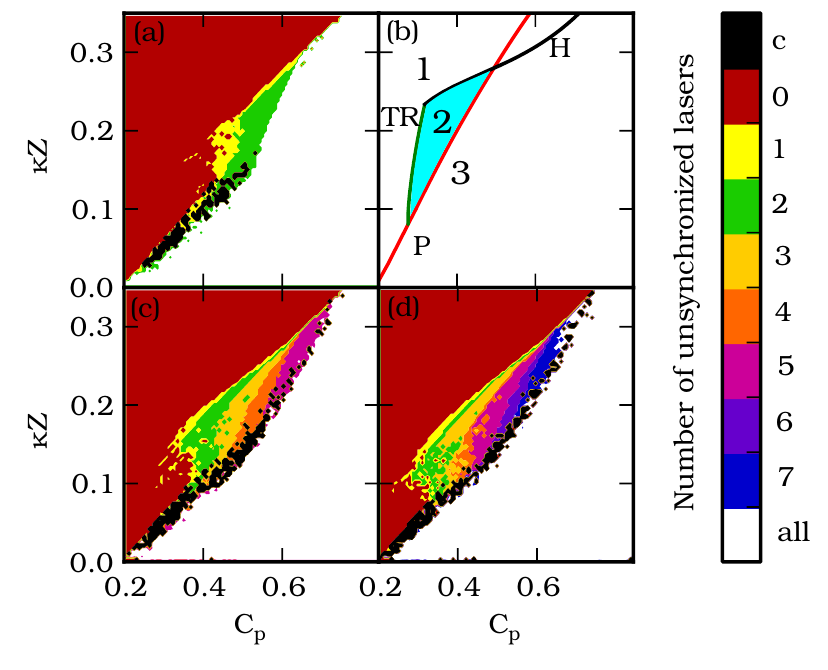}
    \caption{Synchronization regimes obtained by numerical simulation with 
    delay (a),(c),(d) and path continuation without delay (b) for different system sizes: (a) $Z=4$, (b) 2, (c) 8, (d) 
    12 lasers. At the transition from synchronization to desynchronization, partial sychronization occurs for all system 
    sizes (indicated by the color code). Path continuation (b) shows that 
    the regimes coincides with a region of multistability (green, labeled 2) between the 
    synchronous cw state (white, labeled 1) and a bistable periodic solution (white, labeled 3). The 
    regime is bounded by a pitchfork (P) and an addition Hopf (H) 
    and a torus (TR) bifurcation line. Parameters: $\kappa=0.017$, $C_p=0.55$, $T=392$, $\alpha=2.5$, 
        $p=0.23$, $\tau=1$ for (a),(c-d), $\tau=0$ for (b).}				
    \label{fig4}
\end{figure}

Fig.\ref{fig3} shows that the domains of coherence and incoherence can be distinguished 
by amplitude, phase, and inversion. This is different from classical chimera states, where 
the coherence-incoherence pattern is prominently manifested in the phase variable.
Here we find that the lasers show chaotic temporal dynamics for phase, 
inversion, and amplitude in both the coherent and incoherent domains. Regarding the stability of these 
coherence-incoherence patterns, we find that they persist even for long simulation times. These characteristics are similiar to 
amplitude-mediated chimera states that were recently reported in Stuart-Landau oscillators (complex-valued Hopf normal forms) with nonlocal \cite{SET13} and global \cite{SET14} coupling. These states typically show chaotic dynamics and coherence-incoherence patterns for both amplitude and phase, and emerge under random initial conditions. It is remarkable that in contrast to the amplitude-mediated chimeras in the Stuart-Landau system, the chimeras in laser networks also form in 
very small networks. We have already pointed out above that the extension of the regime of full synchronization 
scales with the effective coupling strength $\kappa Z$. It is of interest if 
the regime of partial synchronization scales in the same way 
with the system size $Z$. In Fig.\ref{fig4}, we compare the regime of 
partial synchronization in the $(C_p, \kappa Z$) plane for different system sizes $Z$. 
It can be seen that neither the location nor the 
size change significantly with the number of lasers $Z$. We can thus 
conclude that the system size is not a criterion for the formation of 
these chimera states.

The chimera states found here are thus a stunning counterexamples for the common belief that chimera states emerge under specially prepared initial conditions under nonlocal coupling in large networks, since they (i) form under random initial conditions, (ii) for global coupling, and (iii) even for small numbers of coupled lasers. This requires a critical re-assessment of the question of neceassary criteria for the formation of these novel chimera states in laser networks. As we have already established, the laser network 
combines the concept of amplitude-phase coupling and feedback. Furthermore, the Lang-Kobayashi system describes two intertwined dynamical systems: The complex electrical field $E$ in the laser cavity and the inversion $N$ of electrons in the gain medium. Both systems have unique timescales that are summarized in the time constant $T$. 
Finally, self-feedback and optical injection tend to induce 
multistability in semiconductor lasers 
\cite{LUE11b,OTT14,LUE13,WIE02,ERN10b}. To gain a deeper understanding 
of those stable chimera states, we investigate the dependence upon all those three features in the following. 

\begin{figure}[t!]
    \centering
  \includegraphics[width=0.4\textwidth]{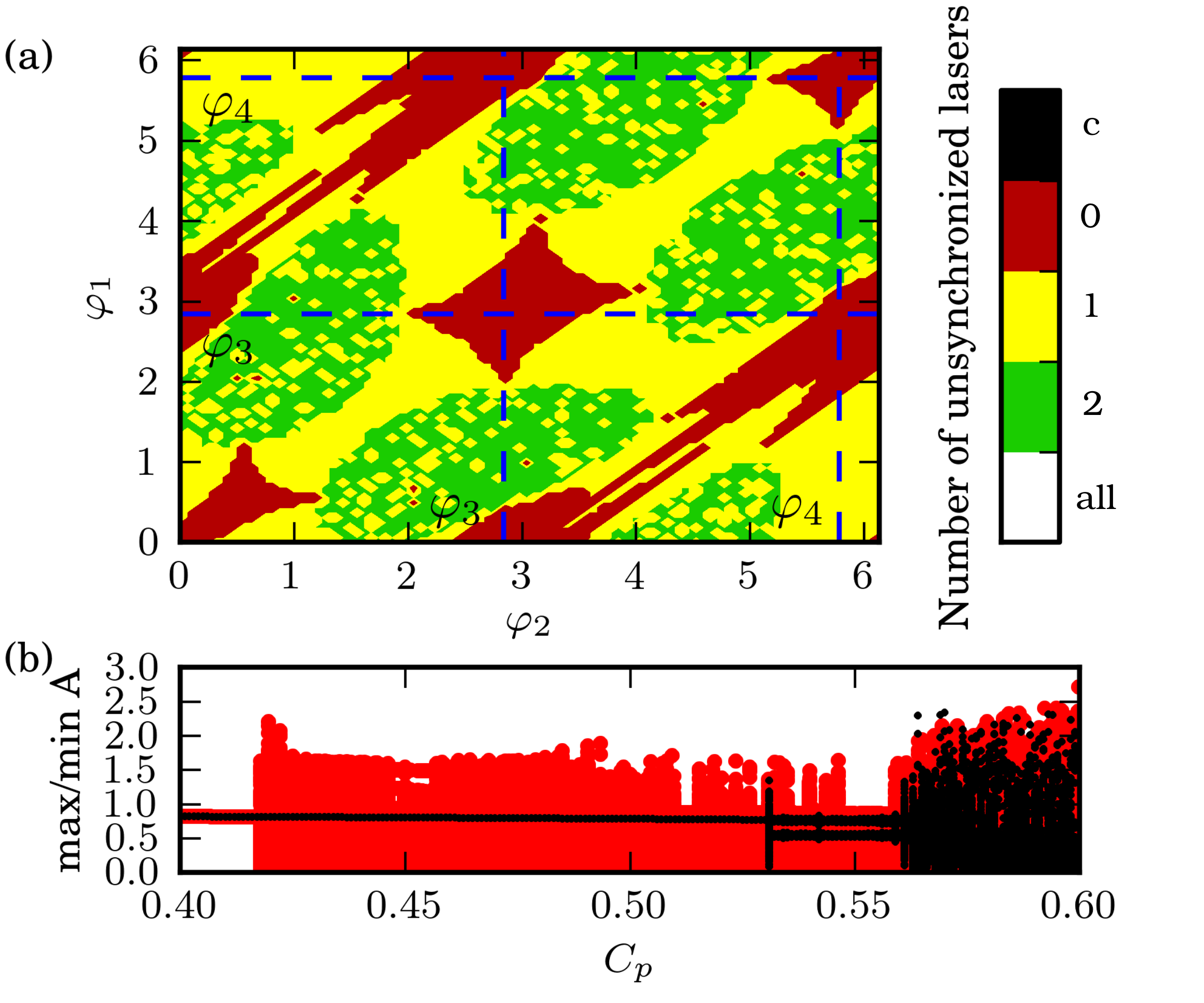}
    \caption{(a): Basins of attraction for synchronization states of 
    fig.\ref{fig4} for $Z=4$, $C_p=0.55$, $\kappa Z=0.2$. in 
    ($\varphi_1, \varphi_2$)-projection.
    Initial phases for laser 3 and 4 are fixed at $\varphi_3=2.839$ and 
    $\varphi_4=5.784$ while $\varphi_1$ and $\varphi_2$ are 
    varied. (b): Linesweep of $C_p$ for fixed $\kappa Z=0.2$ for two 
    different sets of initial phases. In the chimera region, the 
    synchronous CW solution (black) coexists with the partially synchronized 
    chimera state (red). Parameters: $T=392$, $\tau=1$, $p=0.23$, 
    $\alpha=2.5$.}
    \label{fig5}
\end{figure}

To investigate if multistability is important for the formation of chimeras, we 
study the bifurcation structure using numerical path continuation for $Z=2$ lasers. For continuation of the periodic 
solution, we use a previously proposed decomposition \cite{CLE14}. The results are shown in fig.\ref{fig4}(b). 
We find that the 2-laser model exhibits multistability in the region 2 (green)
close to the pitchfork bifurcation line. The boundaries of this region are a pitchfork (P) bifurcation on one side and a 
Hopf (H) and a torus (TR) bifurcation on the other side. Inside the region of multistability 
the synchronous CW state as well as two additional periodic 
solutions are stable. The size and position of region 2 is very 
similiar to the regimes of partial synchronization in Figs.\ref{fig4}(a),(c),(d). This region of 
multistability is also unique and cannot be found in any other part of the 
parameter space. To verify that this bifurcation structure applies also to 
larger systems with time delay, we probe the dynamical behaviour in a linescan of the 
($C_p, \kappa Z$) parameter space for fixed $\kappa$ across the region of multistability in a 4-laser-network for different 
initial conditions. The results in Fig.\ref{fig5}(b) show that there is 
multistability inside the region of partial synchronization depending 
of the choice of initial phases. 
The basin of attraction for 
the 4-laser network inside the multistable region for $C_p=0.5$ is 
mapped in Fig.~\ref{fig5}(a). While the initial states for two lasers remain fixed at $\varphi_3=2.839$ and $\varphi_4=5.784$, the initial phases of the other lasers are varied. If the 
phases are chosen close to in-phase synchronization as initial 
conditions, the synchronous CW state is asymptotically reached from inside its basin of attraction. If 
the phases are randomly distributed along the complex unit circle, the 
chimera state is asymptotically reached.  We can thus conclude that multistability is an important feature associated with the 
formation of the chimera states. Similiar results were 
also found for systems of globally coupled phase 
oscillators \cite{YEL14}. It is well established that strong 
amplitude-phase coupling is able to induce multistability in single 
lasers with optical feedback \cite{OTT12,GLO12}. We observe the same 
behaviour for laser networks. From the path continuation of the 2-laser model, we find that the regime of 
multistability shown in Fig.\ref{fig4}(b) shrinks with decreasing $\alpha$ 
and finally vanishes. The same can be observed for the exemplary 
4-laser network where the chimera regime shrinks and vanishes with 
decreasing $\alpha$. Amplitude-phase coupling is thus crucially connected with the formation of the 
chimera states as it is necessary for the emergence of multistability. 

Another important ingredient of the laser model is the local dynamics 
which is not governed simply by a phase and an amplitude variable, 
like, e.g., the normal form models for Hopf bifurcation (Stuart-Landau 
oscillator) \cite{ZAK14,SCH14a} or the paradigmatic FitzHugh-Nagumo model 
\cite{OME13}, but has higher dimensionality due to an additional 
dynamical variable, the inversion, see Eq.~\eqref{N}, which defines class B 
lasers. In contrast, class A lasers have only two dynamical variables, i.e., amplitude and phase of the complex electric field, and they are characterized by two strongly separated time-scales of the slow electric field amplitude (or optical intensity) and the fast electronic inversion.
This poses the question if the inclusion of the dynamic inversion $N$ 
is necessary for partial synchronization and chimera states. Adiabatic 
elimination of the inversion via the assumption of a quasi-steady state 
$N_{i}^*=\frac{p-|E_i|^2}{1+2|E_i|^2}$ from Eqs.~\eqref{E}, \eqref{N} 
results in: 
\begin{equation}
    \frac{dE_i}{dt}=(1+i\alpha)E_i N_{i}^* + e^{iC_p}\kappa 
    \sum_{j=1}^{Z}e^{iC_p}E_j(t-\tau)
    \label{E_adia}
\end{equation}
Simulating Eq.~\eqref{E_adia} we find that the 
lasers are fully synchronized in the whole parameter space. This is corroborated by the observation that no region of 
multistability can be observed through path continuation of the 2-laser model either. We can thus conclude that 
three dynamic degrees of freedom of the individual lasers are neceassry for the formation of the chimera state in a 
coupled laser system.

In conclusion, we have found chimera states in networks of 
semiconductor class B lasers.
The chimera states differ from classical chimera states in that they show the coherence-incoherence patterns not only in the phase or in the amplitude, but simultaneously in amplitude, phase, and inversion of the laser. In this respect, our model class is essentially different from previously studied model classes like Kuramoto phase oscillators, Stuart-Landau amplitude-phase oscillators, or FitzHugh-Nagumo relaxation oscillators, which have only one or two dynamic degrees of freedom. Furthermore, the dynamics of both the coherent and incoherent parts is purely chaotic in time. 
These intriguing chimera states refute the common paradigm of classical chimera states in several aspects:
First, the stability of the chimera states does not depend on the system size. In fact, we have observed stable chimera states in very 
small networks of only four coupled lasers. Second, global coupling is sufficient to generate these chimera state.
Third, we have found chimera states not only for specially prepared, but for random initial conditions.

We have been able to link the regime where the chimera states emerge to a regime of multistability. The emergence of the multistable regime is correlated with the strong effect of amplitude-phase coupling in the 
semiconductor laser array and vanishes for weak amplitude-phase coupling described by a very small linewidth 
enhancement factor $\alpha$. Furthermore, it was found that the 
lifetime of electrons and photons have to be of the same order of 
magnitude. 
This singles out class B lasers as promising candidates for laser 
chimeras, as opposed to class A lasers.
Experimental realization of these hybrid states seems possible, since 
only global coupling is necessary, which can be easily realized via
a common mirror, and small arrays are sufficient to generate stable chimera states. 

This work was supported by the DFG in the framework of the SFB910.

\bibliographystyle{prsty-fullauthor}

\begin{thebibliography}{10}

\bibitem{HAK83}
H. Haken, {\em {Synergetics, An Introduction}}, 3 ed. (Springer, Berlin, 1983).

\bibitem{PIK01}
A. Pikovsky, M.~G. Rosenblum, and J. Kurths, {\em Synchronization, A Universal
  Concept in Nonlinear Sciences} (Cambridge University Press, Cambridge, 2001).

\bibitem{BOC02}
S. Boccaletti, J. Kurths, G. Osipov, D.~L. Valladares, and C.~S. Zhou, Phys.
  Rep. {\bf 366},  1  (2002).

\bibitem{KUR02a}
Y. Kuramoto and D. Battogtokh, Nonlin. Phen. in Complex Sys. {\bf 5},  380
  (2002).

\bibitem{ABR04}
D.~M. Abrams and S.~H. Strogatz, Phys.~Rev.~Lett. {\bf 93},  174102  (2004).

\bibitem{MOT10}
A.~E. Motter, Nature Physics {\bf 6},  164  (2010).

\bibitem{PAN14}
M.~J. Panaggio and D.~M. Abrams, arXiv:1403.6204  (2014).

\bibitem{HAG12}
A.~M. Hagerstrom, T.~E. Murphy, R. Roy, P. H{\"o}vel, I. Omelchenko, and E.
  Sch{\"o}ll, Nature Physics {\bf 8},  658  (2012).

\bibitem{VIK14}
E.~A. Viktorov, T. Habruseva, S.~P. Hegarty, G. Huyet, and B. Kelleher, Phys.
  Rev. Lett. {\bf 112},  224101  (2014).

\bibitem{TIN12}
M.~R. Tinsley, S. Nkomo, and K. Showalter, Nature Physics {\bf 8},  662
  (2012).

\bibitem{MAR13}
E.~A. Martens, S. Thutupalli, A. Fourri{\`e}re, and O. Hallatschek, Proc. Nat.
  Acad. Sciences {\bf 110},  10563  (2013).

\bibitem{KAP14}
T. Kapitaniak, P. Kuzma, J. Wojewoda, K. Czolczynski, and Y.~L. Maistrenko,
  Scientific Reports {\bf 4},  6379  (2014).

\bibitem{LAR13}
L. Larger, B. Penkovsky, and Y.~L. Maistrenko, Phys. Rev. Lett. {\bf 111},
  054103  (2013).

\bibitem{WIC13}
M. Wickramasinghe and I.~Z. Kiss, PLoS ONE {\bf 8},  e80586  (2013).

\bibitem{SCH14a}
L. Schmidt, K. Sch{\"o}nleber, K. Krischer, and V. Garcia-Morales, Chaos {\bf
  24},  013102  (2014).

\bibitem{OME12a}
O.~E. Omel'chenko, M. Wolfrum, S. Yanchuk, Y.~L. Maistrenko, and O. Sudakov,
  Phys. Rev.~E {\bf 85},  036210  (2012).

\bibitem{PAN13}
M.~J. Panaggio and D.~M. Abrams, Phys. Rev. Lett. {\bf 110},  094102  (2013).

\bibitem{SHI04}
S.-i. Shima and Y. Kuramoto, Phys. Rev.~E {\bf 69},  036213  (2004).

\bibitem{gu13}
C. Gu, G. St-Yves, and J. Davidsen, Phys.~Rev.~Lett. {\bf 111},  134101
  (2013).

\bibitem{OME13}
I. Omelchenko, O.~E. Omel'chenko, P. H{\"o}vel, and E. Sch{\"o}ll, Phys. Rev.
  Lett. {\bf 110},  224101  (2013).

\bibitem{LAI10}
C.~R. Laing, Phys. Rev. E {\bf 81},  066221  (2010).

\bibitem{BOR10}
G. Bordyugov, A. Pikovsky, and M.~G. Rosenblum, Phys. Rev. E {\bf 82},  035205
  (2010).

\bibitem{SET14}
G.~C. Sethia and A. Sen, Phys. Rev. Lett. {\bf 112},  144101  (2014).

\bibitem{ZAK14}
A. Zakharova, M. Kapeller, and E. Sch{\"o}ll, Phys.~Rev.~Lett. {\bf 112},
  154101  (2014).

\bibitem{LAZ14}
N. Lazarides, G. Neofotistos, and G. Tsironis, arXiv {\bf 1408.6072},
  (2014).

\bibitem{RAT00}
N.~C. Rattenborg, C.~J. Amlaner, and S.~L. Lima, Neurosci. Biobehav. Rev. {\bf
  24},  817  (2000).

\bibitem{LAI01}
C.~R. Laing and C.~C. Chow, Neural Computation {\bf 13},  1473  (2001).

\bibitem{SAK06a}
H. Sakaguchi, Phys. Rev.~E {\bf 73},  031907  (2006).

\bibitem{MOT13a}
A.~E. Motter, S.~A. Myers, M. Anghel, and T. Nishikawa, Nature Physics {\bf 9},
   191  (2013).

\bibitem{GON14}
J.~C. Gonzalez-Avella, M.~G. Cosenza, and M.~S. Miguel, Physica~A {\bf 399},
  24  (2014).

\bibitem{ASH14}
P. Ashwin and O. Burylko, arXiv {\bf 1407.8070},    (2014),
  http://arxiv.org/abs/1407.8070.

\bibitem{BAN14}
T. Banerjee, arXiv {\bf 1409.7895},    (2014).

\bibitem{YEL14}
A. Yeldesbay, A. Pikovsky, and M. Rosenblum, Phys. Rev. Lett. {\bf 112},
  144103  (2014).

\bibitem{XIE14}
J. Xie, E. Knobloch, and H.-C. Kao, Phys. Rev.~E {\bf 90},  022919  (2014).

\bibitem{VIE14}
D. Viennot and L. Aubourg, arXiv: {\bf 1408.4585v1},    (2014).

\bibitem{LAR14}
L. Larger, B. Penkovsky, and Y.~L. Maistrenko, arXiv {\bf 1411.4483},
  (2014).

\bibitem{HEI01b}
T. Heil, I. Fischer, W. Els{\"a}{\ss}er, J. Mulet, and C.~R. Mirasso,
  Phys.~Rev.~Lett. {\bf 86},  795  (2001).

\bibitem{WUE05a}
H.~J. W{\"u}nsche, S. Bauer, J. Kreissl, O. Ushakov, N. Korneyev, F.
  Henneberger, E. Wille, H. Erzgr{\"a}ber, M. Peil, W. Els{\"a}{\ss}er, and I.
  Fischer, Phys.~Rev.~Lett. {\bf 94},  163901  (2005).

\bibitem{HIC11}
K. Hicke, O. {D'Huys}, V. Flunkert, E. Sch{\"o}ll, J. Danckaert, and I.
  Fischer, Phys. Rev.~E {\bf 83},  056211  (2011).

\bibitem{FLU11b}
V. Flunkert and E. Sch{\"o}ll, New. J. Phys. {\bf 14},  033039  (2012).

\bibitem{SOR12}
M.~C. Soriano, G. Van~der Sande, I. Fischer, and C.~R. Mirasso, Phys. Rev.
  Lett. {\bf 108},  134101  (2012).

\bibitem{SOR13}
M.~C. Soriano, J. Garc{\'i}a-Ojalvo, C.~R. Mirasso, and I. Fischer,
  Rev.~Mod.~Phys. {\bf 85},  421  (2013).

\bibitem{ZAK10a}
A. Zakharova, T. Vadivasova, V. Anishchenko, A. Koseska, and J. Kurths, Phys.
  Rev. E {\bf 81},  011106  (2010).

\bibitem{KLA10}
S.~H.~L. Klapp and S. Hess, Phys. Rev.~E {\bf 81},  051711  (2010).

\bibitem{LIN12b}
B. Lingnau, K. L{\"u}dge, W.~W. Chow, and E. Sch{\"o}ll, Phys. Rev. E {\bf 86},
   065201(R)  (2012).

\bibitem{LIN14}
B. Lingnau, W.~W. Chow, and K. L{\"u}dge, Opt. Express {\bf 22},  4867  (2014).

\bibitem{MEL06}
S. Melnik, G. Huyet, and A.~V. Uskov, Opt. Express {\bf 14},  2950  (2006).

\bibitem{GRI08}
F. Grillot, N.~A. Naderi, M. Pochet, C.~Y. Lin, and L.~F. Lester, Appl. Phys.
  Lett. {\bf 93},  191108  (2008).

\bibitem{SCH07}
{\em Handbook of Chaos Control}, edited by E. Sch{\"o}ll and H.~G. Schuster
  (Wiley-VCH, Weinheim, 2008), second completely revised and enlarged edition.

\bibitem{FLU13}
V. Flunkert, I. Fischer, and E. Sch{\"o}ll, Theme Issue of Phil. Trans.~R.
  Soc.~A {\bf 371},  20120465  (2013).

\bibitem{LAN80b}
R. Lang and K. Kobayashi, IEEE J. Quantum Electron. {\bf 16},  347  (1980).

\bibitem{ROZ75}
N.~N. Rozanov, Sov. J. Quant. Electron. {\bf 4},  1191  (1975).

\bibitem{ERN10b}
T. Erneux and P. Glorieux, {\em Laser Dynamics} (Cambridge University Press,
  UK, 2010).

\bibitem{KOZ00}
G. Kozyreff, A.~G. Vladimirov, and P. Mandel, Phys.~Rev.~Lett. {\bf 85},  3809
  (2000).

\bibitem{KOZ01}
G. Kozyreff, A.~G. Vladimirov, and P. Mandel, Phys. Rev. E {\bf 64},  1
  (2001).

\bibitem{VLA03}
A.~G. Vladimirov, G. Kozyreff, and P. Mandel, EPL (Europhysics Letters) {\bf
  61},  613  (2003).

\bibitem{YAN04c}
S. Yanchuk, K.~R. Schneider, and L. Recke, Phys. Rev. E {\bf 69},  056221
  (2004).

\bibitem{SET13}
G.~C. Sethia, A. Sen, and G.~L. Johnston, Phys. Rev. E {\bf 88},  042917
  (2013).

\bibitem{LUE11b}
K. L{\"u}dge,  in {\em Nonlinear Laser Dynamics - From Quantum Dots to
  Cryptography}, edited by K. L{\"u}dge ({Wiley-VCH}, Weinheim, 2012).

\bibitem{OTT14}
C. Otto, {\em Dynamics of Quantum Dot Lasers -- Effects of Optical Feedback and
  External Optical Injection}, {\em Springer Theses} (Springer, Heidelberg,
  2014).

\bibitem{LUE13}
K. L{\"u}dge, B. Lingnau, C. Otto, and E. Sch{\"o}ll, Nonlinear Phenom. Complex
  Syst. {\bf 15},  350  (2012).

\bibitem{WIE02}
S. Wieczorek, B. Krauskopf, and D. Lenstra, Phys.~Rev.~Lett. {\bf 88},  063901
  (2002).

\bibitem{CLE14}
E. Clerkin, S. O'Brien, and A. Amann, Phys. Rev. E {\bf 89},  032919  (2014).

\bibitem{OTT12}
C. Otto, B. Globisch, K. L{\"u}dge, E. Sch{\"o}ll, and T. Erneux, Int. J.
  Bifurcation Chaos {\bf 22},  1250246  (2012).

\bibitem{GLO12}
B. Globisch, C. Otto, E. Sch{\"o}ll, and K. L{\"u}dge, Phys. Rev.~E {\bf 86},
  046201  (2012).

\end{thebibliography}

\end{document}